\documentclass{cta-author}

\usepackage{algorithm}
\usepackage{algorithmic}
\usepackage[title]{appendix}

\usepackage{cleveref}
\usepackage{url}
\def\LB{\left(}         
\def\RB{\right)}        

\newfont{\bbb}{msbm10 scaled 500}

\newfont{\bb}{msbm10 scaled 1100}

\newcommand{\kv}{{\bf k}}

\newcommand{\Km}{{\bf K}}

\newcommand{\epsilonv}{\hbox{\boldmath$\epsilon$}}

\newcommand{\phiv}{\hbox{\boldmath$\phi$}}

\newcommand{\thetav}{\hbox{\boldmath$\theta$}}

\renewcommand{\arg}{{\hbox{arg}}}

\usepackage{bbm}
\usepackage{subcaption}
\usepackage{tabularx}
\usepackage{tikz}
\usepackage{multirow}
\usepackage{verbatim}
\usepackage{color}
\usepackage{mathrsfs}
\usepackage{float}
\usepackage{bm}
\usepackage{booktabs}
\usepackage{diagbox}
\include{defn}

\newcommand{\beqa}{\begin{eqnarray}}
\newcommand{\eeqa}{\end{eqnarray}}

\begin{document}


\title{Data-Driven Detection and Identification of IoT-Enabled Load-Altering Attacks in Power Grids}

\author{\au{Subhash Lakshminarayana$^{1}$}, \au{Saurav Sthapit$^{2}$}, \au{Hamidreza Jahangir$^{1}$}, \au{Carsten Maple$^{2}$}, \au{H. Vincent Poor$^{3}$}}
\address{\add{1}{School of Engineering, University of Warwick, UK}
\add{2}{Warwick Manufacturing Group, University of Warwick, UK}
\add{3}{Department of Electrical and Computer Engineering, Princeton University, Princeton, NJ 08544, USA}
\email{\{subhash.lakshminarayana, saurav.sthapit,     hamidreza.jahangir, cm\}@warwick.ac.uk, poor@princeton.edu}}

\begin{abstract}
{Advances in edge computing are powering the development and deployment of Internet of Things (IoT) systems to provide advanced services and resource efficiency. However, large-scale IoT-based load-altering attacks (LAAs) can seriously impact power grid operations, such as destabilising the grid’s control loops. Timely detection and identification of any compromised nodes are essential to minimise the adverse effects of these attacks on power grid operations. In this work, two data-driven algorithms are proposed to detect and identify compromised nodes and the attack parameters of the LAAs. The first method, based on the Sparse Identification of Nonlinear Dynamics (SINDy) approach, adopts a sparse regression framework to identify attack parameters that best describe the observed dynamics. The second method, based on physics-informed neural networks (PINN), employs neural networks to infer the attack parameters from the measurements. Both algorithms are presented utilising edge computing for deployment over decentralised architectures. Extensive simulations are performed on IEEE 6-,14- and 39-bus systems to verify the effectiveness of the proposed methods. Numerical results confirm that the proposed algorithms outperform existing approaches, such as those based on unscented Kalman filter, support vector machines (SVM), and neural networks (NN),  and effectively detect and identify locations of attack in a timely manner.}
\end{abstract}


\maketitle

\section{Introduction}
Critical infrastructures (such as power grids, transportation systems, and nuclear plants) are subject to unprecedented cyber attacks against their system operations \cite{Karnouskos2011, Ukraine2016:Analysis, LakshWiSec2018}.
In particular, power grids are attractive targets since their disruption can have severe social and economic consequences.
The growing integration of IoT-enabled smart-home appliances at the consumer end, such as Wi-Fi-enabled air conditioners, battery storage systems, plug-in-electric vehicles, and so forth, presents a new threat to power grid operations.
{These intelligent appliances are already widely prevalent since they provide convenience to consumers and improve energy efficiency. In the year 2020, it was assessed that globally, there are 30 billion  IoT-enabled devices \cite{CR1}.}
However, due to their poor security features, \cite{Fernandes2016HomeApp, maple2017security}, they may become convenient entry points for malicious parties to gain access to the system. {Thus cyber security can be the main impediment in the widespread adoption of IoT-enabled electrical appliances.}

Cyber security of power grids has received widespread attention in recent years. A significant body of work is dedicated to utility-side cyber attacks and the associated SCADA system security. In particular, false data injection (FDI) attacks against power system state estimation have received significant attention \cite{Liu2009, LakshRMT2021, LakshGT2021}. Different from the aforementioned works, this paper considers cyber attacks that target the end-user consumer devices \cite{HamedLAA2011}, and their impact on power grid operations. {Unlike utility-side and SCADA assets, end-user devices (such as IoT-smart home appliances) can be compromised by adversarial units due to their large numbers, and inefficient monitoring }. Thus, cyber threats targetting demand-side attacks pose unique challenges to power grid operators.


Sudden and abrupt manipulation of the power grid demand due to large-scale cyber attacks against IoT-smart-home appliances (e.g., a Botnet-type attack) can potentially disrupt the balance between supply and demand and lead to severe effects on power grid operations.
Recent works have quantified their impact on power grid operations. In particular, \cite{Dabrowski2017, Soltan2018} showed that LAAs could lead to unsafe frequency excursions, subsequently triggering generator trips and resulting in large-scale blackouts. Moreover, such attacks can also be used to increase the operational cost of the power grid \cite{Soltan2018}. Subsequent work \cite{HuangUSENIX2019} showed that existing protection mechanisms such as under-frequency load shedding (UFLS) are capable of preventing large-scale blackouts. Nevertheless, LAAs can still partition bulk power systems and/or a controlled load shedding event.   Demand-side cyber attacks can also propagate from the distribution grid to the transmission grid, which can lead to an attack impact over a larger geographical area \cite{Dvorking2017}. Reference \cite{Ospina2021} studied the feasibility of LAAs under different power grid loading conditions.

\emph{Static LAAs} leading to a sudden one-time manipulation of the demand (as proposed in \cite{Dabrowski2017, Soltan2018, Dvorking2017, HuangUSENIX2019}) cannot destabilize the power grid's frequency control loop. A more severe attack known as the \emph{dynamic LAAs} was introduced in \cite{AminiLAA2018}, which showed that if the attacker manipulates the load over a period of time in accordance with the fluctuations of the grid frequencies, the resulting attack can destabilize the frequency control loop, which can subsequently trigger cascading failures. 
An analytical method to understand the impact of static/dynamic LAAs using the theory of second-order dynamical systems was presented in \cite{LakshIoT2021}, which was used to find the locations of nodes from which an attacker can launch the most impactful LAAs. Subsequent work investigated techniques to enhance the resilience of power grids to LAAs by improving the security features of a fraction of smart loads \cite{AminiLAA2018, LakshIoT2021}, and finding generator operating points that ensure that no transmission lines are overloaded after any potential LAAs \cite{SoltanTNSE2020}.

Despite the growing number of works on analyzing the impact and improving the resilience of power grids against LAAs, there is relatively less work on detecting and identifying such attacks in real-time, which is the main focus of this paper. Following the onset of an LAA, swift detection and identification of the compromised nodes are critical to minimizing the adverse effects and service disruptions. { In this work, we focus specifically on detecting and identifying LAAs by  real-time  monitoring  and  analysis  of  the  grid’s  physical  signals  such  as  the  voltages, frequencies,  etc. The growing deployment of phasor-measurement units (PMUs) provide fine-grained monitoring of these signals. The key advantage of this approach is that grid operators can
leverage existing sensing infrastructure in the power grid (e.g., advanced metering infrastructure, phasor
measurement units, etc.), and no major device upgrades to IoT-enabled electrical appliances (e.g., enabling encryption/ device remote attestation) would be necessary. Such an approach is particularly relevant in the context of IoT-based attacks due to the massive number of devices and lack of a unified approach to implementing security standards.}

The task is challenging, however, since well-designed LAAs can remain undetected by traditional pattern recognition algorithms applied to the load consumption data \cite{AminiSGC2015}. Amini et al. \cite{AminiSGC2015} showed that LAAs can however be detected by analyzing the system measurements (e.g., load, frequency, etc.) in the frequency domain since LAAs can result in new frequency components that are non-existent in the original data. The method was subsequently extended to identifying the location of the compromised nodes using an optimization-based framework in \cite{AminiIdentification2019}. An unscented Kalman filter (UKF) based approach to detect and identify the attack parameters was proposed in \cite{IzbickiACC2017} in which the attack parameters were augmented to the power grid's state variables and estimated jointly.

However, a major limitation of existing work (such as \cite{AminiSGC2015, AminiIdentification2019}) is that the formulation used is specific to a linear power system model. While the theory of UKF proposed in \cite{IzbickiACC2017} is applicable for non-linear systems, when the power grid dimensions grow, the number of attack parameters to be estimated (which in turn depends on the different combinations of nodes that can be attacked as well as the attacker's chosen parameters) scales much faster than the number of system state measurements. Thus, the UKF approach is not scalable to power grids of large dimensions since the estimation problem becomes under-specified \cite{IzbickiACC2017}. Moreover, our results show that the UKF approach cannot track the attack parameters accurately when the system is highly dynamic.

To overcome the aforementioned issues, this paper adopts state-of-the-art data-driven methods for dynamical systems in order to detect and identify LAAs using real-time monitoring of the power grid's physical signals, such as frequency/phase angle measurements monitored by PMUs. The application is based on the observation is that LAAs lead to a change in the parameters of the power system's dynamical model \cite{AminiLAA2018}. Thus, the attack detection/identification involves solving the following question: \emph{given a series of frequency/phase angle observations over time, what are attack the parameters that best describe the observed data?} We model the problem as one of \emph{data-driven discovery} of power grid's non-linear dynamical equations \cite{BruntonSINDy2016, RaissiPINN2018}. 

We apply two state-of-the-art data-driven methods to solve the aforementioned problem, namely (i) a Sparse Regression (SR) algorithm and (ii) a physics-informed neural network (PINN) algorithm \cite{RaissiPINN2018}. 
The SR algorithm is motivated by the Sparse Identification of Nonlinear Dynamics (SINDy) approach proposed in \cite{BruntonSINDy2016}. It applies sparse regression to determine the fewest attack parameters that best describe the observed measurements. In the context of attack detection/identification, the rationale for seeking a sparse solution is that in a real-world scenario, the attacker can likely monitor the frequency fluctuations and manipulate the load at only a few nodes in the power grid. In particular, we use the least absolute shrinkage and selection operator (LASSO) method to obtain a sparse solution to the LAA detection/identification problem.  
The PINN method on the other hand uses deep neural networks (DNNs) to approximate both the state of the power grid (voltage phase angles/frequency) as well as the unknown non-linear dynamics. It treats the DNN weights as well as the attack parameters as variables to obtain the parameters that best describe the observed dynamics and the mechanisms that govern the evolution of the observed measurements. In both the algorithms, the solution to the attack parameter estimation problem will automatically determine which load buses are compromised by the attacker and those that are not. Furthermore, we present a methodology for the implementation of each method, particularly showing the implementation utilizing edge computing deployed over decentralized architectures, requiring limited information exchange between the neighboring nodes.

{
We note that there is a growing body of work that applies ML to detect FDI attacks against power grids.
Despite their effectiveness, they cannot be applied directly to the problem at hand, since (i) we are interested in both detecting the attacks as well as identifying the attack locations. Existing works on FDI attack detection (e.g., \cite{OzayML2016, HeDL2017, YingSSDL2021}) apply binary classification solutions, i.e., classify whether the attack has occurred or not; they do not localize the attack. Identifying the attack location requires a multi-class classification approach. For the problem at hand, such an approach will have exponential complexity, since the defender must consider every combination of victim nodes that the attacker can target. To overcome this issue, the proposed SR and PINN algorithms leverage the knowledge of the power grid's ``physical model" to infer the attack parameters with a small amount of training data. (ii) ML techniques applied to the FDI attack detection problem only consider the steady-state operation of power grids.
In contrast, our work considers ML algorithms that consider the power grid dynamics and applies data-driven methods for dynamical systems.  }

We conduct extensive simulations using benchmark IEEE bus systems to evaluate the performance of these algorithms in detecting and identifying LAAs. {Our results show that both the SR and the PINN algorithms are more effective in estimating the attack parameters as compared to the other benchmark approaches, including support vector machines (SVM), NN, and the UKF approach, especially in systems that exhibit fast dynamics.} Furthermore, both the algorithms can estimate the attack parameters within a short observation time window. However, the PINN algorithm does not perform well on systems exhibiting slow, and oscillatory dynamics since the training may get stuck in local minima. On the other hand, the SR algorithm performs reliably well under all system conditions.

{To summarize, the main contributions of this work are as follows:
\begin{itemize}
\item Proposing novel data-driven approaches (SR and PINN) to detect and localize IoT-based load-altering attacks against power grids.
\item Estimating the static/ dynamic attack parameters by monitoring the frequency/phase angle dynamics over an observation time window. 
\item Verifying the effectiveness of the proposed methods on slow oscillatory and fast dynamics networks.
\item Comparing the performance of the proposed methods with benchmark techniques, including SVM, NN and UKF.
\end{itemize}
}

The rest of the paper is organized as follows. Section \ref{sec:Prelim} introduces the system model; Section~\ref{sec:Algos} describes the SR and the PINN algorithms. Section~\ref{sec:Implement} includes discussion on practical aspects as well as a decentralized implementation of the proposed algorithms. 
Section~\ref{sec:Sims} describes the simulation results and Section~\ref{sec:Conc} concludes. The simulation parameters are listed in the Appendix. 


\section{Preliminaries}
\subsection{System Model}
\label{sec:Prelim}
We consider a power grid consisting of a set of $\mathcal{N} = \{1,\dots,N \}$ buses. For each node $i \in \mathcal{N},$ the set of neighboring nodes is denoted by $\mathcal{N}_i.$ The buses are divided into generator buses $\mathcal{N}_G$ and load buses $\mathcal{N}_L$ and $\mathcal{N} = \mathcal{N}_G \cup \mathcal{N}_L$. The power grid dynamic model is given by  \cite{kundur1994}:
\begin{align}
  \dot{\delta}_i & = \omega_i,  i \in \mathcal{N}_G \label{eqn:dyn1} \\
 M_i \dot{\omega}_i & = - D_i \omega_i - K^P_i  \omega_i - K^I_i  \delta_i  - P^F_i,  \ i \in \mathcal{N}_G, \label{eqn:dyn2}\\
D_i \dot{\delta}_i &=   - P^{L}_i  - P^F_i, \ i \in \mathcal{N}_L, \label{eqn:dyn3}
\end{align}
where $P^F_i = \sum_{j \in \mathcal{N}} B_{i,j} \sin (\delta_{ij}).$ In the above, $\delta_i$ is the phase angle deviation at bus $i \in \mathcal{N},$ 
$\delta_{ij} = \delta_i - \delta_j, \ i,j \in \mathcal{N}$ and
$\omega_i$ denotes the rotor frequency deviation at the generator buses $i\in \mathcal{N}_G.$ The generator inertia coefficient at bus $i \in \mathcal{N}_G$ is denoted by $M_i.$ The damping coefficients at generator/load buses is denoted by $D_i$ $i \in \mathcal{N}$. 
$B_{i,j}$ is the suscpetance of line $i,j.$ Finally, $P^L_i $ denotes the load at bus $i \in \mathcal{N}_L$.

We denote $\omega_{\text{nom}}$ as the grid's nominal frequency, e.g., $50~$Hz in Europe or  $60~$Hz in North America. For safe operations, the frequency must be maintained within the safety limits. We denote $\omega_{\text{max}}$ as the maximum permissible frequency deviation for system safety. Thus, $| \omega_{\text{nom}} - \omega_i| \leq \omega_{\text{max}}, \forall i \in \mathcal{G}.$ We note that in steady state, $\dot{\omega}_i = 0, \forall i \in \mathcal{G}.$

\subsection{Load Altering Attacks}
{
IoT-enabled end-user appliances often have poor security features. Several such vulnerabilities have been discovered in commercial IoT high-wattage appliances. For instance, it has been shown that Mitsubishi ACs are prone to XML external entity injection (XXE) vulnerabilities \cite{mitsibushi1,controllerEW50A}.
They can be exploited to launch denial-of-service (DoS) or privilege escalation (unauthorized access) attacks. In particular, the latter can be exploited by an external entity (such as an attacker) to change the operational settings of ACs (e.g., switch ON/OFF or change the power settings). Similarly, other IoT power appliances such as photovoltaic (PV) inverters  \cite{solarinverter} and electric vehicle (EV) chargers \cite{Acharya2020} are also known to have cyber vulnerabilities.

Botnet-type attacks against IoT devices, such as the Mirai attack that controlled up to 600,000 devices, have been well-documented \cite{Mirai2017}. If such attacks are launched against IoT-enabled electrical appliances, they can significantly disrupt power grid operations. Assuming that a commercial AC has a power rating of $2~$kW, the exploitation of tens of thousands of such devices can lead to several megawatts of load alteration, significantly disrupting the balance between supply and demand.
In this work, we do not explicitly model how an attacker can launch Botnet-type attacks against the power grid. Rather, inspired by the aforementioned vulnerabilities, we focus on the impact of such attacks on power grid operations and the corresponding detection/identification mechanisms. In the following, we present a model of LAAs in the context of power grid control loops described in \eqref{eqn:dyn1}-\eqref{eqn:dyn3}.
}

Under IoT-based LAAs, the attacker manipulates the system load by synchronously switching on or off a large number of high-wattage devices. Assume that the demand at the load buses consists of two components
$P^L_i = P^{LS}_i + P^{LV}_i,$ where $P^{LS}_i$ denotes the secure part of the load (i.e., load that cannot be altered) and $ P^{LV}_i$ denotes the vulnerable part of the load at node $i \in \mathcal{N}_l$. 
We denote the set of victim nodes by $\mathcal{N}_v (\subseteq \mathcal{N}_L),$ and $N_v = |\mathcal{N}_v|,$ which are the subset of load buses at which the attacker can manipulate the load.  
The system load under LAAs is given by
\begin{align}
P^L_i = -\sum^S_{j = 1} K^{L}_{i,k_j} \omega_{k_j}  + \epsilon^L_i + P^{LS}_i \label{eqn:load_attack}
\end{align}
Herein, $\epsilon^L_i$ is a step-change in the load introduced by the attacker (static LAA component). Note $\epsilon^L_i = 0$ if $i \notin \mathcal{N}_v.$ The component $-\sum_{k_j \in \mathcal{S}} K^{L}_{i,k_j} \omega_{k_s}$ is the dynamic LAA component. Note that to inject the dynamic LAA, the attacker must monitor the frequency fluctuations at a subset of the buses in the system,
$\omega_{k_j}, k_j \in \mathcal{S}.$ Herein,  $\mathcal{S} (\subseteq \mathcal{N}),$ denotes the set of \emph{sensing} buses, i.e., buses at which the attacker can monitor the frequency fluctuations, and $k_j$ denotes the index of the $j^{\text{th}}$ sensing bus and $S = |\mathcal{S}|$. $K^{L}_{i,k_j}$ denotes attack controller gain values corresponding to the attack at load bus $i \in \mathcal{N}_L$ that follows the frequency at bus $k_j \in \mathcal{S}.$ Note that ${K}^{L}_{i,k_s} \geq 0,$ and therefore, the dynamic LAA component increases the system load when the frequency falls below the setpoint and vice versa; this has the opposite effect of arresting the frequency deviation. Furthermore, we have $\epsilon^L_i  \leq P^{LV}_i$ and 
\begin{align}
 \sum^S_{j = 1} {K}^{L}_{i,k_j} \omega^{\max}   \leq (P^{LV}_i - \epsilon^L_i)/2 , \forall i \in \mathcal{V}. \label{eqn:Attack_lim}
\end{align}
The above limit on the dynamic LAA component can be explained as follows. 
The left-hand side of \eqref{eqn:Attack_lim} is the maximum load value that must be altered by the attacker at victim bus $i \in \mathcal{V}$ before the frequency at the sensor bus $k_j$ exceeds the  safety limit $\omega^{\max}$. This must be less than the amount of vulnerable load $P^{LV}_i - \epsilon^L_i$ (after removing the static LAA component). Note that the amount of load that can be compromised under dynamic LAA must allow for both over and under frequency fluctuations before the system frequency exceeds  $\omega^{\max}$ (see  \cite{AminiLAA2018}). Thus, the right-hand side of \eqref{eqn:Attack_lim} is divided by $2.$
Under LAA described in \eqref{eqn:load_attack}, the power grid dynamics noted in  \eqref{eqn:dyn1}-\eqref{eqn:dyn3}  becomes,
\begin{align}
  \dot{\delta}_i & = \omega_i,  i \in \mathcal{N}_G \label{eqn:dyn_attack1} \\
 M_i \dot{\omega}_i & = - D_i \omega_i - K^P_i  \omega_i - K^I_i  \delta_i   - P^F_{i},  \ i \in \mathcal{N}_G, \label{eqn:dyn_attack2}\\
D_i \dot{\delta}_i = &  \sum^S_{j = 1} K^{L}_{i,k_j} \omega_{k_j}  -  \epsilon^L_i - P^{LS}_i - P^F_{i}, \ i \in \mathcal{N}_L, \label{eqn:dyn3_LAA}
\end{align}
Equations \eqref{eqn:dyn_attack1}, \eqref{eqn:dyn_attack2} and \eqref{eqn:dyn3_LAA} jointly describe the power grid dynamics under LAAs. 

{
\subsection{Impact of LAAs}
To illustrate the attack impact, we show the phase angle and frequency dynamics for the IEEE-6 bus system under LAAs in Fig.~\ref{fig:Attack_Grid}.

{\bf Static LAAs:} First, we consider static LAAs only, i.e., when $\epsilon^L_i \neq 0,$ whereas $K^L_{i,k_j} = 0$ (dynamic component). Note that static LAAs only act as an initial perturbation on the system. Small perturbations (corresponding to natural load fluctuations in the system) lead to minor frequency fluctuations, and the generator's control loops act quickly to counter the frequency deviation. However, LAAs that cause large-scale fluctuation in the system frequency can lead to large frequency fluctuations, leading to unsafe frequency excursions. An example of frequency fluctuations following a static LAA of 0.5~pu (base load of $100~$MW) is plotted shown in Fig.~\ref{fig:Attack_Grid}(a). Note also from Fig.~\ref{fig:Attack_Grid}(a) that small-scale load changes only cause minor frequency deviations.

{\bf Dynamic LAAs:} As compared to static LAAs, dynamic LAAs, i.e., attacks with non-zero values of $K^L_{i,k_j}$ changes the dynamic model of the system, which can be noted by comparing equations \eqref{eqn:dyn3} and \eqref{eqn:dyn3_LAA}. This can alter the system design considerations and potentially destabilize the grid's frequency control loop. In Fig.~\ref{fig:Attack_Grid}(b), we plot the frequency dynamics for different values of $K^L_{i,k_j}$ for IEEE-6 bus system. As evident, the frequency control loop destabilizes for large values of attack controller gain (see \cite{LakshIoT2021} for more analysis on this parameter). Moreover, we note that different values the attack controller gain generate different patterns of frequency dynamics (more discussion is provided in Section~\ref{sec:Sims}). 


\begin{figure}[!t]
\centering
\begin{subfigure}{0.4\textwidth}
\includegraphics[width=1\textwidth]{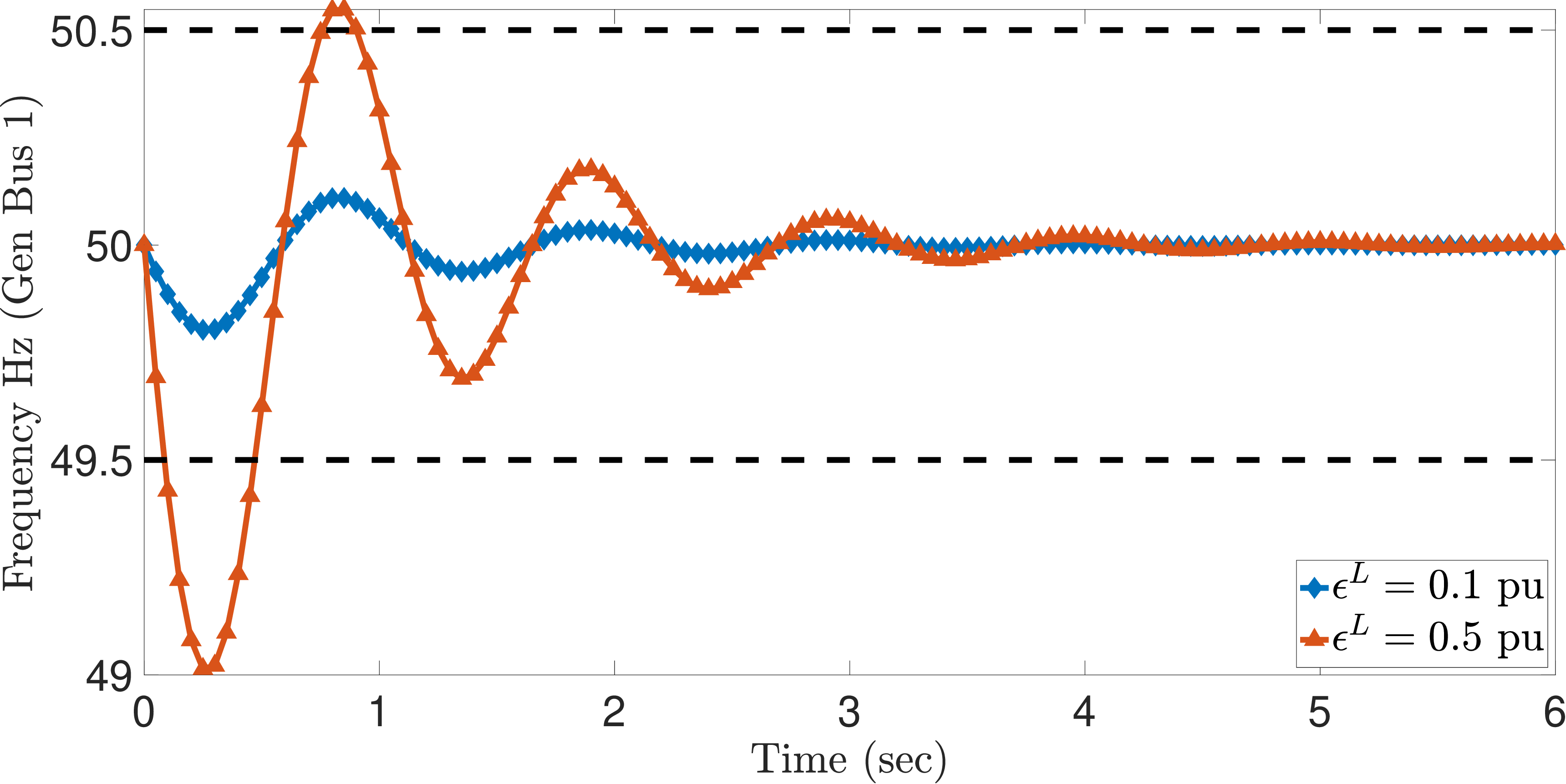}
\caption{}
\end{subfigure}
~
\begin{subfigure}{0.4\textwidth}
\includegraphics[width=1\textwidth]{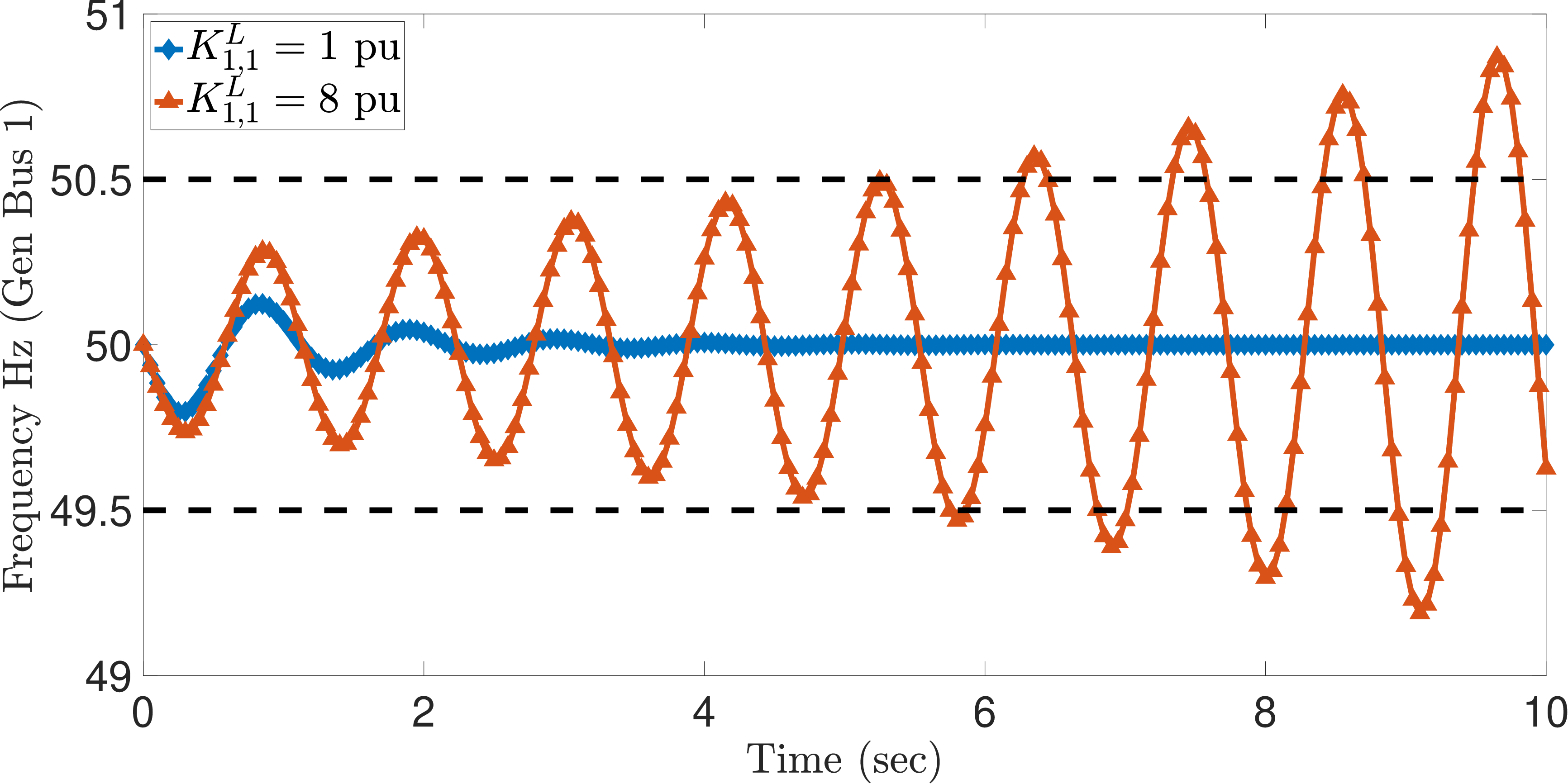}
\caption{}
\end{subfigure}
\caption{{Frequency dynamics under (a) Static LAAs (b) Dyanmic LAAs for IEEE-6 bus system.}}
\label{fig:Attack_Grid}
\vspace{-0.1 cm}
\end{figure}

\section{Data-Driven Techniques for Detecting and Identifying LAAs}
\label{sec:Algos}

\begin{figure}[!t]
\centering
\includegraphics[width=0.48\textwidth]{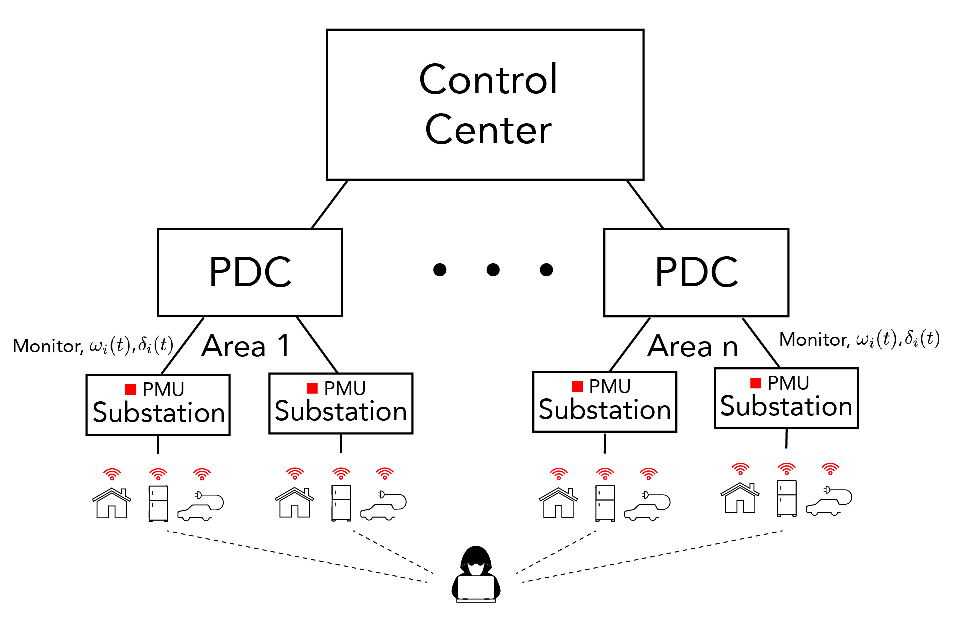}
\caption{Proposed detection framework and integration into wide-area monitoring systems.}
\label{fig:WAMS}
\end{figure}

{
This work focuses on detecting and identifying IoT-based LAAs by real-time monitoring and analysis of the grid’s physical signals. In particular, we assume that the system operator has deployed PMUs in the power grid, which enables them to monitor the voltage phase angles $\{ \delta^{(\tau)}_i \}_{i \in \mathcal{N}, \tau = 1,\dots,T}$ and frequency fluctuations $\{ \dot{\delta}^{(\tau)}_i \}_{i \in \mathcal{N}, \tau = 1,\dots,T}$,  respectively over a period of time. Herein, we assume a slotted time system with $x^{(\tau)}$ as the value of the signal $x$ at time slot $\tau,$ where the slots are sampled at a time interval of $T_s$ and $T$ is the total number of time slots. For instance, according to IEEE/IEC standards, for a $50~$Hz system, the PMU sampling frequency can be between 10 to 100 frames per second. Therefore, $T_s$ is in the range $10-100~$ms \cite{PMU2018}. Attack detection/identification problem involves inferring the parameters $\{{K}^{L}_{i,k_j}\}_{i \in \mathcal{V}, k_j \in \mathcal{S}}$ and $\epsilon^L_i, i \in \mathcal{V}$ by 
by monitoring the power grid dynamics $\{ \delta^{(\tau)}_i \}_{i \in \mathcal{N}, \tau = 1,\dots,T}$ and $\{ \omega^{(\tau)}_i \}_{i \in \mathcal{N}, \tau = 1,\dots,T}.$

The overall implementation of the proposed methodology and its integration into power grid wide-area monitoring systems (WAMS) is shown in Fig.~\ref{fig:WAMS}. IoT-enabled electrical appliances are installed and operated by end-users, who typically lack security hygiene. Moreover, due to their massive numbers, and the lack of a unified approach to implementing security standards, IoT-based electrical appliances are insecure and vulnerable to attacks \cite{Fernandes2016HomeApp, maple2017security}. In contrast, the PMU networks that monitor the power grid's physical signals are deployed by the system operators (e.g., at substations). It is feasible to deploy advanced security solutions to these systems and ensure the confidentiality and integrity of the monitored physical signals. For instance, the IEC 62351 standard provides security recommendations for different power system communication protocols, including the widely used IEC 61850 protocol for PMU/substation communication \cite{IECreview2020}. Thus, we assume that the attacker cannot modify the PMU-monitored signals, making the proposed data-driven attack detection/identification particularly suitable for detecting LAAs. In particular, the proposed approach will localize the buses (i.e., the substations) that are targetted by the attacker under LAAs. We provide further details on the integration of the proposed detection scheme into WAMS in Section~\ref{sec:Implement}.

In particular, we propose physics-informed machine learning algorithms to detect and identify LAAs. We note that the problem at hand can be modelled as supervised learning that learns only based on the training samples and does not rely on any prior knowledge or the laws of physics. Specifically, one can train a machine learning classifier that distinguishes between the system's dynamics under different values of the attack parameters $\{{K}^{L}_{i,k_j}\}_{i \in \mathcal{V}, k_j \in \mathcal{S}}, \epsilon^L_i, i \in \mathcal{V}.$ However, under such an approach, identifying the locations and the magnitude of LAA will be challenging since it will involve a combinatorial search across nodes of the grid that the attacker may target (note that this number can  potentially  be $2^{|\mathcal{V}| \times |\mathcal{S}|}$). Training such a classifier would be computationally complex, which requires a tremendous amount of training data.

On the other hand, if the physical model governing the system dynamics is known, this knowledge can be incorporated into the learning. This approach is beneficial when the inference must be performed with a limited number of samples, which is the case in our problem, where the identification must be completed quickly before the system is damaged. Motivated by these challenges, we propose two state-of-the-art techniques for data-driven identification of dynamical systems to solve the LAA detection/identification problem, namely, the SR algorithm and the PINN algorithm. The details are presented next.


\subsection{LAA Detection and Identification Using SR Approach}
The SR approach, which is based on the SINDy algorithm proposed in \cite{BruntonSINDy2016}, applies sparse regression to find the attack parameters that best represent the observed power grid dynamics. To formalize the SR framework for detecting and identifying LAAs, we define the following additional notations. First, we define $\dot{\theta}_i$ as
\begin{align}
\dot{\theta}^{(\tau)}_i = \dot{\delta}^{(\tau)}_i + \frac{1}{D_i} \LB P^{LS}_i + \sum_{j \in \mathcal{N}} B_{i,j} \sin (\delta^{(\tau)}_{ij})   \RB. \label{eqn:rearr}
\end{align}
Using \eqref{eqn:rearr}, a discretized version of the load bus dynamics in \eqref{eqn:dyn3_LAA} over $\tau = 1,2,\dots,T$ time slots can be rewritten as
\begin{align}
\underbrace{\begin{bmatrix}
\dot{\theta}^{(1)}_i \\
    \vdots  \\
\dot{\theta}^{(T)}_i \\ 
\end{bmatrix}}_{\dot{\thetav}_i}  = \underbrace{\frac{1}{D_i} \begin{bmatrix}
\omega^{(1)}_{k_1},\dots,\omega^{(1)}_{k_S},-1 \\
    \vdots  \\
\omega^{(T)}_{k_1},\dots,\omega^{(T)}_{k_S},-1 \\ 
\end{bmatrix}}_{\Omega } \underbrace{\begin{bmatrix}
K^L_{i,k_1} \\
    \vdots  \\
K^L_{i,k_s}  \\
\epsilon_i
\end{bmatrix}}_{\kv^L_i}, i \in \mathcal{N}_L. \label{eqn:Regression_full}
\end{align}
Equation \eqref{eqn:Regression_full} represents the relationship between the attack parameters $\kv^L_i$ and the load bus dynamics at bus $i \in \mathcal{N}_L.$
Note that the system operator can compute the elements of  $\dot{\thetav}_i$ and $\Omega$ using the frequency and phase angle measurements obtained from PMU. 
Thus, the system operator can estimate the attack parameters $\kv^L_i$ by solving \eqref{eqn:Regression_full}.

At this stage, several comments are in order. First,
\eqref{eqn:Regression_full} is a system of $T$ linear equations with $S+1$ unknown variables. In general $T \gg S+1$, since PMUs have a high sampling frequency (e.g., $50$ measurement samples per second \cite{PMU2018}). Second, the measurements in $\dot{\thetav}_i$ and $\Omega$ are corrupted by measurement noises. Third, the value of $\kv^L_i$ that solves \eqref{eqn:Regression_full} may not be unique. For instance, there may be multiple attack vectors that result in similar dynamics. Thus, we seek to find the sparsest attack vector $\kv^L_i$ that best explains the dynamics observed in $\dot{\thetav}_i$ and $\Omega$. The rationale is that in a real-world scenario, the attacker can likely monitor the frequency fluctuations and manipulate the load at only a few nodes in the power grid.
To promote sparsity, we use the LASSO algorithm, which applies an L1-regularization term to promote sparsity. The optimization problem corresponding the LASSO algorithm is given by \cite{LassoRef1996}
\begin{align}
   \widehat{\kv}^{L}_i = \arg \min_{\kv^{L}_i} || \Omega \kv^{L}_i  - \dot{{\thetav}}_i  ||^2 + \lambda || \kv^{L}_i ||_1, \ i = 1 \dots, |\mathcal{V}|. \label{eqn:sparse_reg}
\end{align}
Herein, $|| \kv^{L}_i ||_1$ represents a penalty term that and $\lambda >0$ is a scaling parameter.  {The overall SR approach is described in Algorithm~1.}

\begin{algorithm}[htb]
	\caption{{Sparse Regression Algorithm}}
    {\textbf{Input:} Phase angle data $\{ \delta^{(\tau)}_i \}_{i \in \mathcal{N}, \tau = 1,\dots,T}$ and frequency data $\{ \omega^{(\tau)}_i \}_{i \in \mathcal{N}, \tau = 1,\dots,T}$} \\
    {\textbf{Output:} Estimates of the static ($\widehat{\epsilon}^L_i, i = 1,\dots,N_L$) and dynamic ($\widehat{K}^L_{i,k_j},  i = 1,\dots,N_L, j = 1,\dots,S$) LAA parameters} \\
	\begin{algorithmic}[1]
	      \STATE {Monitor phase angle $\{ \delta^{(\tau)}_i \}_{i \in \mathcal{N}, \tau = 1,\dots,T}$ and frequency $\{ \omega^{(\tau)}_i \}_{i \in \mathcal{N}, \tau = 1,\dots,T}$ data. 
	      \STATE From the phase angle data and frequency data, compute $\{ \dot{\theta}^{(\tau)}_i \}_{i \in \mathcal{N}, \tau = 1,\dots,T}$ as in \eqref{eqn:rearr}.
	      \STATE Construct the vector and the vector $\dot{\thetav}$ and the matrix $\Omega$ from the phase angle and frequency measurements.
	      \STATE Using $\dot{\thetav}$ and $\Omega$, solve for $\widehat{\kv}^L_i$ using the LASSO algorithm.
	      \STATE Output $\widehat{\kv}^L_i, i = 1,\dots,|\mathcal{N}_L|.$}
	\end{algorithmic}
\end{algorithm}

\subsection{Identification Based on Physics-Informed Neural Networks}
We now describe the PINN framework  \cite{RaissiPINN2018, Lagaris1998} applied to detecting/identifying LAAs. 
PINN incorporates the knowledge of the physical model guiding the observed dynamics to train a neural network with limited training samples. As in \cite{RaissiPINN2018}, we incorporate the underlying differential equations as an additional loss function to solve two different types of problems. The first problem is the forward problem, which uses NN to obtain a solution to the differential equations, i.e., in our context, the signals $\delta_i(t)$ and $\omega_i(t)$ that solve \eqref{eqn:dyn_attack1} , \eqref{eqn:dyn_attack2} and \eqref{eqn:dyn3_LAA} under LAAs. The second problem that is of more interest to our work is the inverse problem, in which, given the measurements and the structure of the differential equations, estimate the unknown parameters $\{{K}^{L}_{i,k_j}\}_{i \in \mathcal{V}, k_j \in \mathcal{S}}$ and $\epsilon^L_i, i \in \mathcal{V}$ that best describe the observed data.

The PINN framework accomplishes these two tasks by defining appropriate loss functions and training a NN to minimize them. The overall PINN framework is shown in Fig.~\ref{fig:pinn_arch} and 
we present the details next. First, let $\widehat{\omega}_i (t,\phiv)$ and $\widehat{\delta}_i (t,\phiv)$
be the NN approximations of ${\omega_i} (t)$ and ${\delta_i} (t)$
as shown in Fig.~\ref{fig:pinn_arch}. Herein,  $\phiv$ represents the weights of the NN, which are trained to minimize the following losses:

\subsubsection{Mean-square loss}
The first loss term involves the mean square loss between the observed measurements ${\omega_i} (t)$ and ${\delta_i} (t)$ and their NN approximations, i.e., 
\begin{align}
    L_1 (\phiv) =\frac{1}{T}\sum_{\tau=1}^{T} & \Big{(} \sum_{i \in \mathcal{N}_G} (\widehat{\omega}_i(\tau ,\phiv) \nonumber  - {\omega}_i(\tau))^2 \\ & + \sum_{i \in \mathcal{N}} (\widehat{\delta}_i(\tau,\phiv) - {\delta}_i(\tau))^2 \Big{)}.
\end{align}\label{mean_square_loss}
\subsubsection{Physics based Loss}
The system dynamics are incorporated as a second loss function. Specifically, let us define
\begin{align}
 f^{(1)}_i & =  \dot{\widehat{\delta}}_i  - \widehat{\omega}_i,  i \in \mathcal{N}_G \label{eqn:dyn11} \\
f^{(2)}_i & =  M_i \dot{\widehat{\omega}}_i   + D_i \widehat{\omega}_i + K^P_i  \widehat{\omega}_i \nonumber \\ & \qquad + K^I_i   \widehat{\delta}_i  
 + \sum_{j \in \mathcal{N}} B_{i,j} \sin (\widehat{\delta}_{ij}),  \ i \in \mathcal{N}_G, \label{eqn:dyn12}\\
f^{(3)}_i & = D_i \dot{\widehat{\delta}}_i - \sum^S_{j = 1} \widehat{K}^{L}_{i,k_j} \widehat{\omega}_{k_j}  +  \widehat{\epsilon}^L_i  + P^{LS}_i  \nonumber \\ & \qquad + \sum_{j \in \mathcal{N}} B_{i,j} \sin (\widehat{\delta}_{ij}), \ i \in \mathcal{N}_L, \label{eqn:dyn13}
\end{align}
where in the above, we have dropped the notations showing the dependency of $f^{(1)}_i$ and $f^{(2)}_i$ on $(t,\phiv)$ and $f^{(3)}_i$ on $(t, \phiv, \Km^L, \epsilonv^L)$ and simplicity.
In the above, the derivatives are computed using NN's automatic differentiation.
The physics-based loss function is then given by
\begin{align}
    L_2  & (\phiv, \Km^L, \epsilonv^L)   =\frac{1}{T}\sum_{\tau=1}^{T} \Big{(} \sum_{i \in \mathcal{N}_G} (f^{(1)}_i (\tau, \phiv))^2  \nonumber \\ &  +  \sum_{i \in \mathcal{N}_G} (f^{(2)}_i(\tau, \phiv)^2 +  \sum_{i \in \mathcal{N}_L} (f^{(3)}_i(\tau, \phiv, \Km^L, \epsilonv^L)^2 \Big{)}.
\end{align}\label{physics_loss}
Note that the above loss function depends both on the NN weights $\phiv$ and the attack parameters $\Km^L,\epsilonv^L.$

\subsubsection{Sparsity-Promoting Loss}
As noted before, we are interested in sparse solutions to the attack parameter estimation problem. Thus, in the PINN framework, we also add sparsity promoting term in the loss function given by 
\begin{equation}
    L_3(\Km^L) = \alpha \LB || \widehat{\Km}^L ||_1 + || \widehat{\epsilonv}^L ||_1  \RB,
\end{equation}\label{sparsity_promoting_loss}
where, $\alpha$ is the scaling factor and $|| \widehat{\Km}^L ||_1$ and  $|| \widehat{\epsilonv}^L ||_1$ are the 1-norm of the vectors consisting of all the attack parameters. 

The NN is trained using the observed measurements  $\{ \delta^{(\tau)}_i \}_{i \in \mathcal{N}, \tau = 1,\dots,T}$ and $\{ \dot{\delta}^{(\tau)}_i \}_{i \in \mathcal{N}, \tau = 1,\dots,T}$ as follows:
\begin{align*}
  \phiv^*, \widehat{\Km}^L, \widehat{\epsilonv}^L = \arg \min_{\phiv, \Km^L, \epsilonv^L } L_1  (\phiv) + L_2  (\phiv, \Km^L, \epsilonv^L) +  L_3(\Km^L).
\end{align*}
Note that when the sum of these two losses is minimized, 
we ensure three criteria: (i) the output of the NN replicates the observed system dynamics, (ii) $f^{(1)}_i, f^{(2)}_i,$ and $f^{(3)}_i$ are close to zero, which in turn implies that  \eqref{eqn:dyn_attack1}, \eqref{eqn:dyn_attack2} and \eqref{eqn:dyn3_LAA} are satisfied, and (iii) the obtained solution is sparse. This implies that the estimated attack parameters that best fit the observed data. {The overall PINN approach is described in Algorithm~2.}

\begin{figure}
    \centering
    \includegraphics[width=0.48\textwidth]{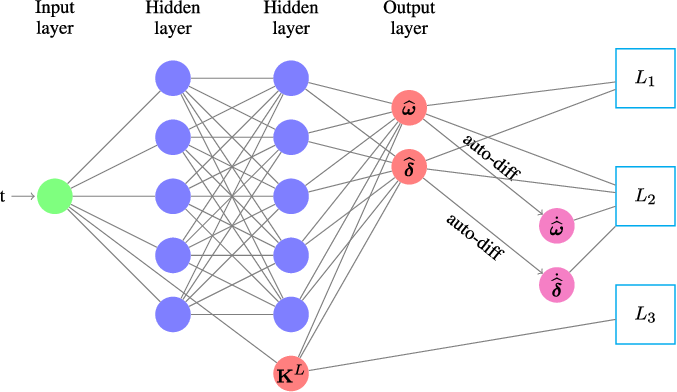}
    \caption{PINN network showing input output layers and the output attack parameters.}
    \label{fig:pinn_arch}
\end{figure}

\begin{algorithm}[htb]
	\caption{{PINN Algorithm}}
    {\textbf{Input:} Phase angle data $\{ \delta^{(\tau)}_i \}_{i \in \mathcal{N}, \tau = 1,\dots,T}$ and frequency data $\{ \omega^{(\tau)}_i \}_{i \in \mathcal{N}, \tau = 1,\dots,T}$ \\
    \textbf{Output:} Estimates of the static ($\widehat{\epsilon}^L_i, i = 1,\dots,N_L$) and dynamic ($\widehat{K}^L_{i,k_j},  i = 1,\dots,N_L, j = 1,\dots,S$) attack parameters \\
    \textbf{Parameters:}  Neural Network Parameters $\phiv$}\\
	\begin{algorithmic}[1]
	     \STATE {Setup Neural network with parameter $\Theta$ and unknown attack parameters (${K}^L_{i,k_j},  i = 1,\dots,N_L, j = 1,\dots,S$) as TensorFlow variables as shown in \cref{fig:pinn_arch}.}
	     \STATE {Use auto-gradient to calculate the derivatives of the phase angle and frequency data.}
	     \STATE {Setup loss functions $ L_1 (\phiv), L_2(\phiv, \Km^L, \epsilonv^L)$ and $L_3(\Km^L)$ defined in \cref{mean_square_loss,physics_loss,sparsity_promoting_loss}.}
	     \STATE {Monitor phase angle $\{ \delta^{(\tau)}_i \}_{i \in \mathcal{N}, \tau = 1,\dots,T}.$ and frequency data $\{ \omega^{(\tau)}_i \}_{i \in \mathcal{N}, \tau = 1,\dots,T}.$ }
	      \STATE { Use the collected data to train the neural network.} 
	      \STATE {Output the attack parameters $\widehat{K}^L_{i,k_j},  i = 1,\dots,N_L, j = 1,\dots,S$}.
	\end{algorithmic}
\end{algorithm}

{
\subsection{Comparison with Black-Box Machine Learning Algorithms}
As noted before, the SR and PINN algorithms are examples of the so-called ``physics-informed" ML algorithms. Alternately, one may also use ``black-box" ML algorithms (e.g., support vector machines or neural networks) that are purely data-driven and do not use the knowledge of the underlying physical models. We now clarify the difference between the two approaches in the context of detection and identification of IoT-enabled LAAs. 

Note that both the SR and the PINN algorithms do not require ``offline" training. The data monitored in real-time can be directly fed as inputs to Algorithms~1 and 2 to generate estimates of the attack parameters. In contrast, the black-box ML algorithms require offline training, which can be very data-intensive, especially for the considered case of IoT-based LAAs. In particular, we must train the ML model with training data corresponding to LAAs at different nodes (note that there can be $2^{|\mathcal{V}|}$ combinations of victim nodes, considering multi-point attacks). Moreover, the attack parameters at the victim nodes can also have different magnitudes. Thus, training ML models considering many combinations will be computationally complex and will require a tremendous amount of training data. In Section~5.4, we provide a numerical comparison of the proposed algorithms with SVM and NN trained using black-box training techniques. We observe that the accuracy of training depends heavily on the range and the amount of the training data. To find more details about SVM and NN, see  \cite{JAHANGIR2020100601}.
}

\section{Discussion and Implementation}
\label{sec:Implement}
In this section, we present a discussion on some practical aspects of the proposed algorithms.
\subsection{{Differentiating Attacks From Natural Load Fluctuations}}
{
 Note that the power system is naturally dynamic, i.e., the frequency/phase angle fluctuations occur due to natural load changes. Thus, the grid operator will need a mechanism to differentiate these natural variations from LAAs. There are two important differences between natural load fluctuations and an LAA: (i) for static LAAs, $\epsilon^L_i$ will have a large value (to cause unsafe frequency excursions, please see Fig.~\ref{fig:Attack_Grid} (a)) (ii) for dynamic LAAs, the attack controller gain parameter $K^L_{i,j}$ will be non-zero and positive. Note that this is designed to work in an opposite manner to the generator's governor control (that arrests deviations), i.e., the dynamic LAA decreases the load when the frequency is increasing (about the setpoint), and LAA increases the load when the frequency is decreasing (below the setpoint) \cite{AminiLAA2018}}.

{In general, the operator can adopt the following procedure to differentiate natural load fluctuations to attacks. (i) Implement the SR or PINN algorithm periodically to infer  $\epsilon^L_i$ and $K^L_{i,j}$ values from the observed data. (ii) If $\widehat{\epsilon}^L_i \neq 0$ and $\widehat{K}^L_{i,j} = 0,$ then, compare $\widehat{\epsilon}^L_i$ against a threshold value, which is computed based on the historical or forecasted value of the load fluctuations. If it exceeds the threshold, raise an attack alert. (iii) If $\widehat{K}^L_{i,j} > 0,$  compare against a threshold (around zero). If it  exceeds the threshold, raise an attack alert.}

\subsection{Implementing the Methods}
This section highlights the implementation aspects of the SR and PINN algorithms. The proposed algorithms can be straightforwardly integrated into existing wide-area monitoring systems (WAMS). The hierarchical structure of the WAMS network is shown in Fig.~\ref{fig:WAMS}. PMUs installed at the substation monitor the phase angle/frequency measurements and send these measurements to a phasor data concentrator (PDC) and subsequently to a control center (CC). The proposed algorithms can be implemented in a centralized (at the CC) or decentralized manner. Since they are not significantly resource-intensive, both the SR and PINN algorithms can be implemented at the edge, either at a local phasor data concentrator (PDC) or at individual substations with limited information exchange between the nodes. This allows efficient response and detection, more control over data flows, and therefore enhanced security, privacy and data handling. We elaborate the method in the following.

Recall that the SR algorithm involves solving \eqref{eqn:sparse_reg} for every victim node $i \in \mathcal{V}$ to determine $\kv_i$. This in turn requires the following signals at each node $i \in \mathcal{V}$:  (i) $ \delta_i$   (ii) $  \{ \delta_{j} \}_{j \in \mathcal{N}_i},$ and (iii) $ \{ \omega_{k_j} \}_{ k_j \in \mathcal{S}}$. Note that (i) can be monitored locally at individual substations, and (ii) only requires information exchange with the neighboring nodes of node $i,$ i.e., $j \in \mathcal{N}_i$. On the other hand, (iii) requires information exchange from all the potential nodes at which the attacker can sense the grid frequency. In practice, due to the attacker's limited capabilities, these nodes are likely to be restricted to nodes that belong to the same control area or those that are connected to the same PDC. Thus, the SR algorithm can be implemented locally with limited information exchange.

The PINN algorithm can be also be implemented in a decentralized manner. In particular, each substation can locally solve the following optimization problem to determine $\kv^L_i$. 

\begin{align}
  \phiv^*_i, \widehat{\kv}^L_i = \arg \min_{\phiv_i, \kv^L_i } L_1  (\phiv_i) + L_2  (\phiv_i, \kv^L_i) +  L_3(\kv^L_i), \label{eqn:PINN_local}
\end{align}

\begin{align}
  \text{where} \  L_1 (\phiv) & =\frac{1}{T}\sum_{\tau=1}^{T}  \Big{(}  \sum_{j \in \{ i \cup \mathcal{N}_i \} } (\widehat{\omega}_i(\tau ,\phiv) - {\omega}_i(m))^2 \nonumber  \\ & + \sum_{j \in \{ i \cup \mathcal{N}_i \} } (\widehat{\delta}_i(\tau,\phiv) - {\delta}_i(m))^2 \Big{)}, \nonumber \\
     L_2 (\phiv) & = (f^{(3)}_i (\phiv_i, \kv^L_i))^2,
     L_3 (\phiv)  =  \alpha || \kv^L_i ||_1. \nonumber
\end{align}
As noted in the case of the SR algorithm, all the signals required to solve \eqref{eqn:PINN_local} can be obtained locally with limited information exchange between nodes of the same control area.

\section{Simulations}
\label{sec:Sims}
In this section, we present simulations to show the effectiveness of the proposed approaches. 

\subsection{Simulation Settings and Methodology}

\subsubsection{Bus Systems} 
 {The proposed algorithms are tested using the IEEE 6-,14- and 39-bus systems. } The power grid topological data is obtained from the MATPOWER simulator.
 For the IEEE 39-bus system, $\mathcal{N}_G = \{30,\dots,39\}, |\mathcal{N}_G| = 10$ and  $\mathcal{N}_L = \{1,\dots,29\}, |\mathcal{N}_L| = 29.$ We assume $\mathcal{S} = \mathcal{N}_G$ and $\mathcal{V} = \mathcal{N}_L.$ Thus, in our simulation setup, the attack parameter identification involves estimating $|\mathcal{N}_G| \times |\mathcal{N}_L| = 290$ values of $K^L_{v,s}, v \in \mathcal{V}, s \in \mathcal{S}.$ {For the IEEE 14-bus system, $\mathcal{N}_G = \{1,2,3,6,8\}, |\mathcal{N}_G| = 5$ and  $|\mathcal{N}_L| = 9.$ Finally,for the IEEE 6-bus system, $\mathcal{N}_G = \{1,2,3\}, |\mathcal{N}_G| = 3$ and  $|\mathcal{N}_L| = 3.$ More details on the simulation parameters used in the case studies are listed in the Appendix.}

\subsubsection{Algorithm Implementation and Settings}
We implement $5$ algorithms in total -- (i) SR (ii) PINN (iii) SVM (iv) NN, and (v) UKF. 
The details are as follows: (i) The SR algorithm is 
implemented using the \emph{Lasso} function in MATLAB. 
(ii) The PINN algorithm is implemented using Tensorflow. We use a multi-layer perceptron for PINN with three hidden layers, each with $50$ neurons. We use automatic differentiation to calculate the gradients. The NN is trained using limited memory Broyden–Fletcher–Goldfarb–Shanno algorithm (LBFGS) optimizer (similar to \cite{RaissiPINN2018}), as it is better suited when the training samples are limited. (iii) { SVM is implemented in MATLAB using Gaussian kernel function.} (iv) { NN is also implemented in MATLAB. For NN, three multi-layer perceptrons with ten neurons for each layer is considered. To avoid over-fitting problems, the cross-validation technique and the ReLU activation function are also implemented.} (v) 
Finally, UKF is implemented using \emph{unscentedKalmanFilter} function in MATLAB.

\subsubsection{Generating Training/Testing Data}

\begin{figure}[!t]
\centering
\includegraphics[width=0.47\textwidth]{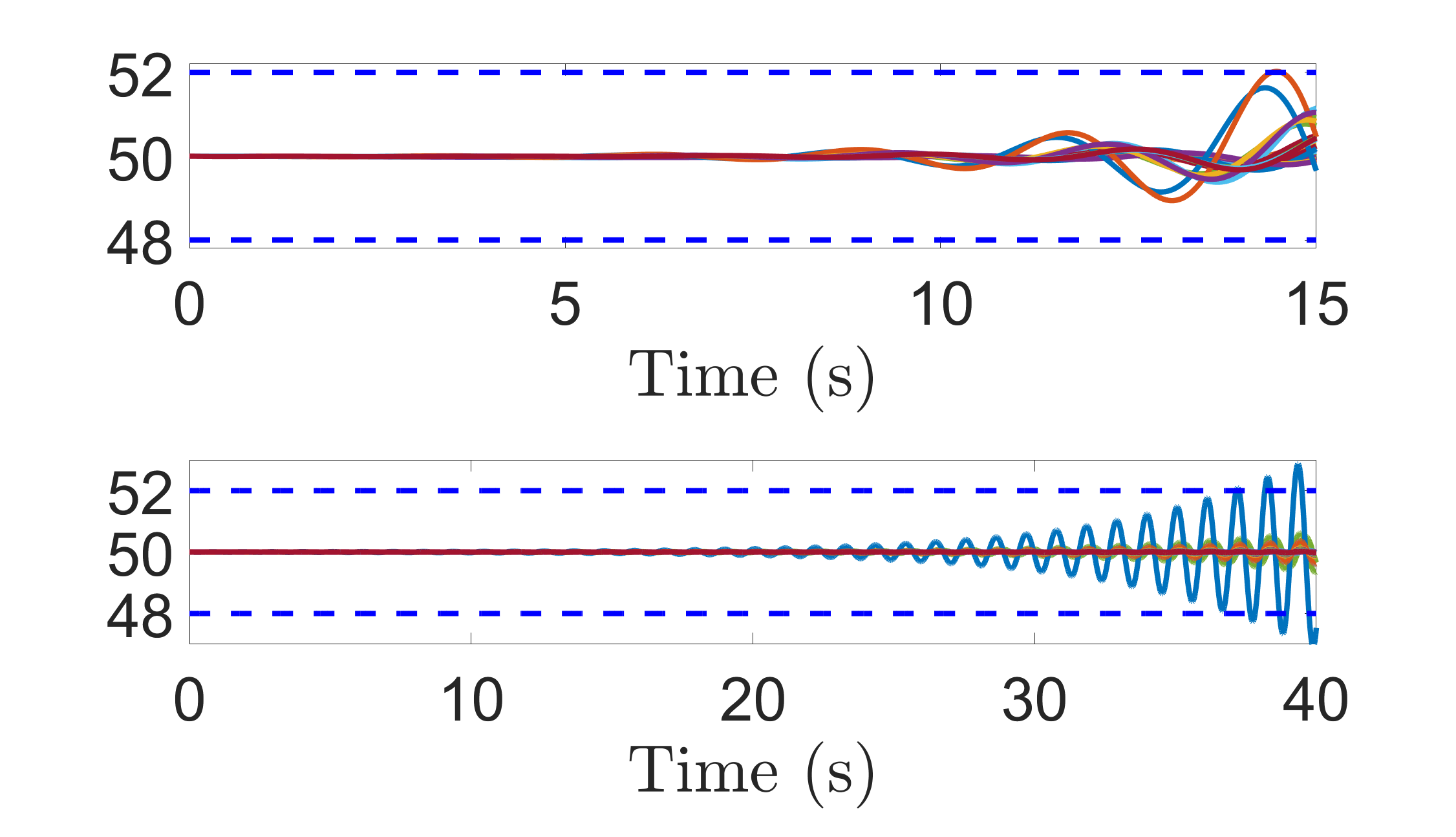}
\caption{Frequency dynamics under DLAAs for the IEEE-39 bus system. Top figure: Fast dynamics (System parameters A) wit $K^L_{19,33} = 18$. Bottom figure: Slow oscillatory dynamics (System parameters B) with $K^L_{19,33} = 25$. The horizontal lines indicate the safety limit of $2~$Hz.}
\label{fig:freq_dyn_plot39bus}
\end{figure}

In order to generate the dataset corresponding to power grid signals under LAA, we simulate the differential equations in \eqref{eqn:dyn_attack1}, \eqref{eqn:dyn_attack2}, and \eqref{eqn:dyn3_LAA} using the \emph{ode-45} function of MATLAB.

For brevity, we only present details from the IEEE-39 bus system (the settings for other bus system can be found in the Appendix). In this bus system, the node pair $(v,s) = (19,33)$ corresponds to the least-effort LAA \cite{LakshIoT2021}. Thus, for single-point attacks, we set  $K^L_{19,33}$ to a non-zero value (to be specified subsequently) and rest of the values of $K^L_{v,s}$ to be zero.
{We consider two sets of generation parameters for each test case as listed in the Appendix. The frequency/phase angle dynamics of the IEEE 39-bus system corresponding to the two sets of parameters are shown in Fig.~\ref{fig:freq_dyn_plot39bus}}. We refer to the parameter set A1 as \emph{fast dynamics}, in which the frequency deviates away from the setpoint of $50~$Hz relatively quickly within a few oscillation cycles. It can be observed that the oscillations are relatively smooth. The attack parameter in this case is set to $K^L_{19,33} = 18.$ 
We refer to parameters set B1 as \emph{slow oscillatory dynamics}, in which we observe several cycles of oscillations before the frequency deviates away from the setpoint of $50~$Hz.
The attack parameter in this case is set to $K^L_{19,33} = 25.$ For both cases, we set $\epsilon_{19} = 0.1$ pu. To generate noisy data (as inputs to the algorithms), we add Gaussian noise to the frequency and the phase angle waveforms (obtained by solving \eqref{eqn:dyn1}-\eqref{eqn:dyn3}), whose standard deviation is set to $0.01~$pu. We assume PMU sampling frequency of $50$ measurements per second. Hence, the number of measurements involved in the training process is $50T,$ where $T$ is the measurement time window (specified subsequently). 

{As explained in Section~3.3, SR and PINN are physics-informed ML algorithms that do not require ``offline training''. Thus, the data generated using the procedure stated above can be directly fed as inputs to the algorithm for inferring the attack parameters. The SVM and NN approaches on the other hand are ``black-box'' ML algorithms that require ``offline training''. In the offline data generation procedure, 2000 samples, based on two features, $\delta_i$ and $\omega_i$, $i\in \mathcal{N}_G$, with 100 time-steps, are generated for each victim bus  $\mathcal{N}_v (\subseteq \mathcal{N}_L)$. To incorporate diversity in the offline-training data, both the location of the sensing bus/victim bus $\mathcal{V}, \mathcal{S} (\subseteq \mathcal{N})$ as well as the 
the dynamic attack parameter $K^L_{v,s}$ are varied between a certain range. 
As noted in Section 3.3, the range of $K^L_{v,s}$ considered in the offline training data is critical to the performance of SVM and NN. To illustrate these effects, we consider two different ranges, (i) $(1.2-23.4)$ and (ii) $(0.8-32.4)$ and discuss the dependence of this parameter in Section 5.4. }

 
\subsubsection{Algorithm Evaluation Metrics} 
Attack detection and identification algorithms in the context of LAAs must have two characteristics (i) the targetted node(s) and the corresponding attack parameter(s) must be identified correctly, and (ii) for the nodes that are not targetted by the attacker, the estimated value of must be close to zero. 
{
Since our objective is not only finding the attack locations, but also determining the attack parameters, we evaluate the algorithm performance in terms of two metrics. The first, given by $$\eta_1 = (K^L_{v,s} - \widehat{K}^L_{v,s})/K^L_{v,s},$$ where $v$ and $s$ are the actual victim and sensing buses respectively (at which the attack is injected to generate the frequency and phase angle dynamics). The second metric is given by $$\eta_2 = \frac{1}{|\mathcal{V}| |\mathcal{S}|}\sum_{v \in \mathcal{V}, s \in \mathcal{S}} (K^L_{v,s} - \widehat{K}^L_{v,s})^2,$$
i.e., the mean-square error of the estimated attack parameters at nodes that are not targetted by the attacker. Note that a low value of $\eta_1$ indicates high precision in estimating the correct attack parameters (true positive), where as a high value indicates that the attack is not detected accurately (false negative). Similarly, a low value of $\eta_2$ indicates that the nodes not targetted by the attacker are not identified as attacked nodes (true negative), where as a high value indicates large number of false positives. }

\subsection{Simulation Results}
We present the simulation results next.
Once again, for brevity, the simulation results from the IEEE-39 bus system are presented as plots and other results from other bus systems are presented in a tabular form.

\subsubsection{Single-Point Attacks}
{We examine the effectiveness of the SR and PINN methods in detecting the DLAAs and compare their performance with the SVM, NN and UKF method.}

{The parameter estimation accuracy $\eta_1$ (boxplots) for the IEEE-39 bus system with fast dynamic parameters is shown in Fig.~\ref{fig:Acc_SR_PINN_SVM_NN} for the SR, PINN, SVM and NN algorithms,
respectively. The results are plotted for different observation time windows. 
The attack parameter estimation is repeated $100$ times, where the measurements differ due to the noise realization. Furthermore, for the range of the dynamic attack parameter in the training data for SVM and NN algorithms, we consider (i) $(1.2-23.4)$ (see Section 5.4 for more discussion on this). }

{{\bf Fast Dynamics:} It can be observed that the SR method can estimate the dynamic attack parameters more accurately compared to PINN, SVM, and NN for fast dynamics before the system dynamics breach the safety limits.} As expected, the attack parameters can be estimated with greater accuracy when the observation time window is larger, due to the availability of more data. 

{{\bf Slow Oscillatory Dynamics: } The SR algorithm achieves high precision for slow-oscillatory dynamics as well. However, we observed that other benchmark algorithms have a large error in inferring the attack parameters in this simulation setting. Accordingly, for this simulation setup, only SR results are presented in Fig.~\ref{fig:Acc_SR}, and other benchmarks, which have more than $50\%$ error are neglected.} 

To further investigate the poor performance PINN, SVM and NN algorithms for slow oscillatory dynamics, we investigate their online training procedure. For brevity, we only present the result for PINN. 
As illustrated in Fig.~\ref{fig:train_epoch}, which presents the overall loss as a function of the training epochs, while in the case of fast dynamics, the error is minimized over several training epochs ($47,000$ epochs) and reaches a very low value, for slow oscillatory dynamics, the training terminates relatively quickly ($640$ epochs) since the loss does not decrease further.
{We conjecture this is because the system with slow dynamics has several oscillatory cycles (see Fig.~\ref{fig:freq_dyn_plot39bus}) and the training process may be potentially stuck at local minima}. In contrast, for the system with fast dynamics, the oscillatory cycles are relatively smooth and conducive to training the NN. {Thus, the PINN algorithm is not suited for attack detection/identification for systems with slow dynamics.} 


\begin{figure}[!t]
\centering
\includegraphics[trim=50 0 0 0,clip, width=0.55\textwidth]{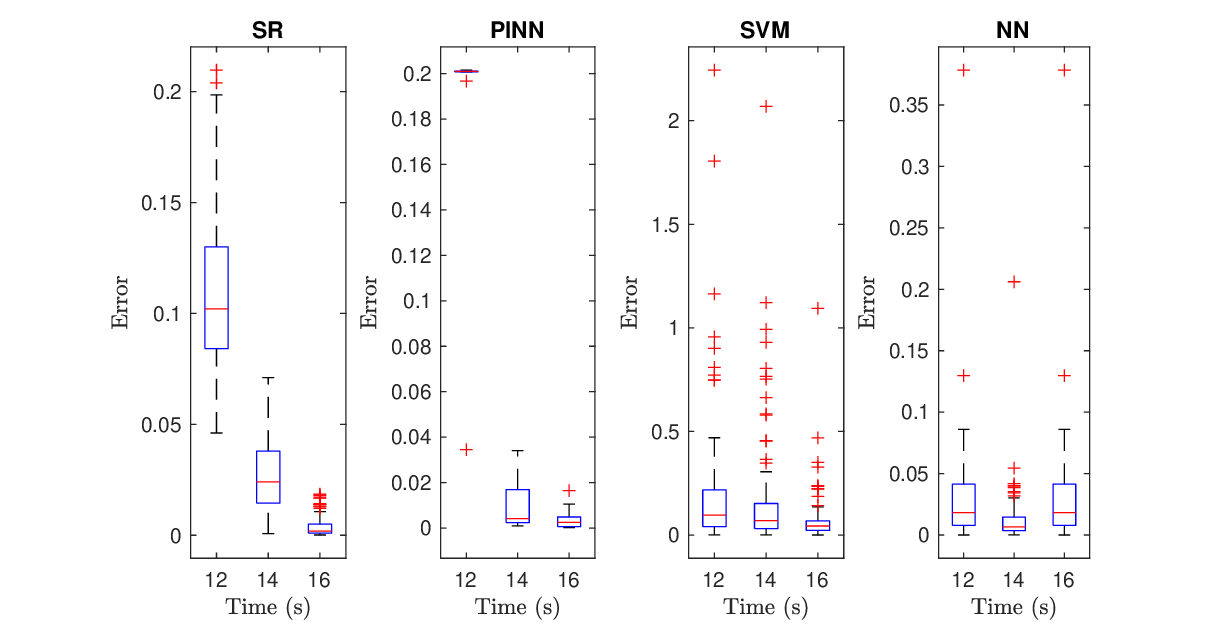}
\caption{{Attack parameter estimation error $\eta_1$ (boxplots) for fast dynamics ($12-16$ s, System parameters A) with $K^L_{19,33} = 18$. }}
\label{fig:Acc_SR_PINN_SVM_NN}
\end{figure}

\begin{figure}[!t]
\centering
\includegraphics[trim=10 0 0 0,clip, width=0.37\textwidth]{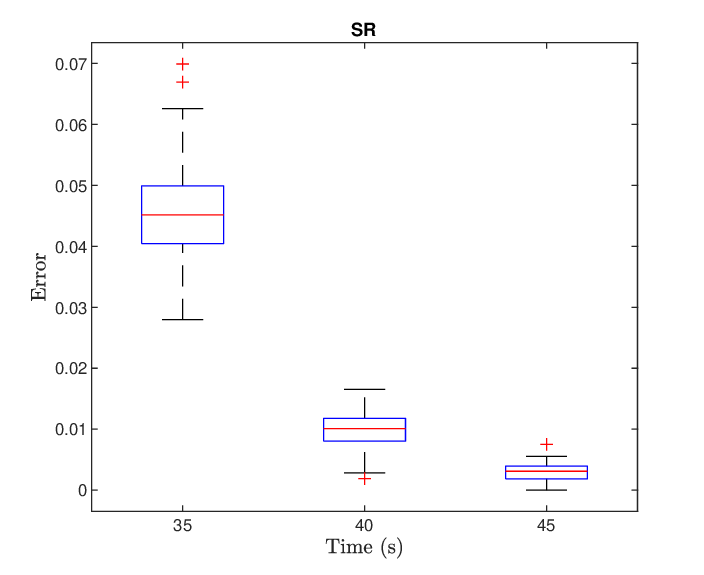}
\caption{{Attack parameter estimation error $\eta_1$ (boxplots) for slow oscillatory dynamics ($30-45$ s, System parameters B) with $K^L_{19,33} = 25$. }}
\label{fig:Acc_SR}
\end{figure}

\begin{figure}
    \centering
    \includegraphics[width=0.4\textwidth]{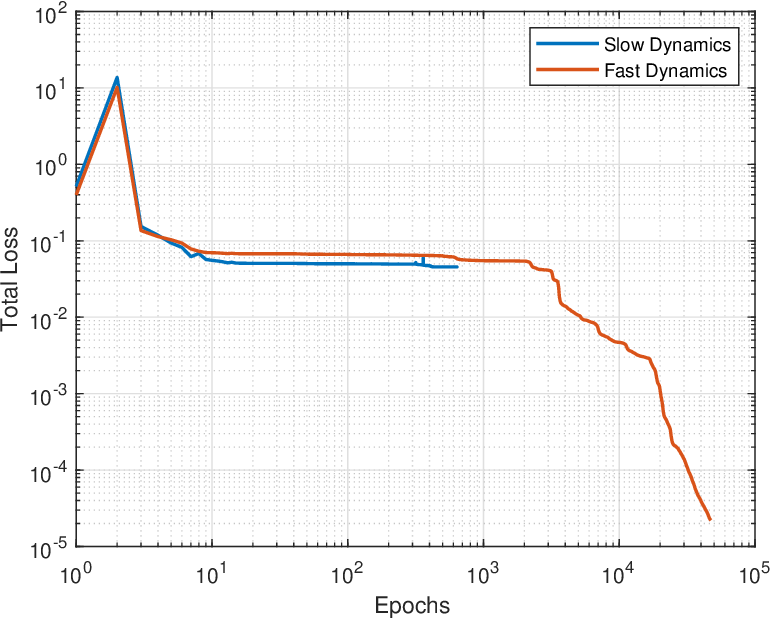}
    \caption{PINN training for the slow and fast dynamics cases showing loss over the training epochs. The  Both X and Y axes are in Logarithmic scale.}
    \label{fig:train_epoch}
\end{figure}

Further, we also enlist the overall mean-square error of attack parameter estimation ($\eta_2$) in Table~\ref{tbl:Accuracy}. {A similar trend as the previous result is observed, i.e., while the SR, PINN, SVM, and NN algorithms achieve good performance for the system with fast. However, for slow dynamics, only the SR algorithm achieves satisfactory performance.}
In particular, the low value of $\eta_2$ indicates that the value of $K^L_{v,s}$ for nodes that are not targetted by the attacker are close to zero. {The error for the PINN, SVM and NN algorithms under slow oscillatory dynamics is very high; hence we do not include them in  Table~\ref{tbl:Accuracy}. }

Finally, we also compare the UKF approach for attack parameter estimation in Table~\ref{tbl:Accuracy} (last column). It can be observed that while UKF method performs well in the case of slow dynamics, it is unable to do so in the case of fast dynamics. This is because UKF is a sequential algorithm, and hence, it is unable to capture the effect of rapid fluctuations in the system dynamics over a relatively short period of time. The poor performance of UKF can also be observed by noting the value of $\eta_1$ in Table \ref{tbl:Accuracy_all_case} (IEEE-39 bus, corresponding to fast dynamics), where we observe an error of $48 \%$ in estimating the value of the attack parameter. 

\begin{table}[!t]
\processtable{Mean-square error of attack parameter estimation ($\eta_2$)  for IEEE-39 bus system with single-point attack.  $K^L_{19,33} = 18$ for fast dynamics (Parameters A) with ($T = 15$ s) and $K^L_{19,33} = 25$ for slow oscillatory dynamics (Parameters B) with ($T = 40$ s).\label{tbl:Accuracy}}
 {\begin{tabular}{c c  c  c  c c  c c} 
 \hline
   & Metric & SR & PINN & {SVM} & {NN } & UKF \\ [0.5ex]
 \hline\hline
{Fast dynamics} & $\eta_2$ & $0.011$   &  $0.08$  &{ $1.350$} & {$0.097$} & $0.44$ \\ \hline 
{Slow oscillatory dynamics} & $\eta_2$ & $0.1$   & $- $   & $- $ & $- $ & $0.08$\\  \hline 
\end{tabular}}{}

\end{table}

\begin{table}[!t]
\processtable{{Simulation results for different case studies under single-points attacks for fast dynamic parameters}\label{tbl:Accuracy_all_case}}
 {\begin{tabular}{c c c  c  c  c c } 
 \hline
   & Metric & SR & PINN &  {SVM} & {NN} & {UKF} \\ [0.5ex] 
 \hline\hline
\multirow{2}{*}{{IEEE 6-Bus System } } & $\eta_1$ & $0.028$ &  $0.07$ & $0.219$ & $0.097$ & $0.47$ \\  
 & $\eta_2$  & $0.038$ & $0.092$ & $1.178$ & $0.808$ & $2.13$ \\ \hline 
\multirow{2}{*}{{IEEE 14-Bus System }} & $\eta_1$  & $0.06$ & $ 0.064 $  & $0.168$ & $0.103$ & $0.49$\\ 
 & $\eta_2$  & $0.0151$ & $ 0.075 $ & $2.355$ & $0.962$ & $3.18$ \\ \hline 
 \multirow{2}{*}{IEEE 39-Bus System } & $\eta_1$   & $0.04$ & $ 0.08 $ & $0.073$ & $0.086$ & $0.48$\\ 
 & $\eta_2$  & $0.011$ & $ 0.085 $ & $1.350$ & $0.097$ & $0.44$\\ \hline 
\end{tabular}}{}

\end{table}

{\subsection{Simulation Results for Different Bus Systems}
To verify the effectiveness of the proposed methods in different bus systems, the results from IEEE 6-,14-, 39- bus systems are presented for fast dynamic parameters. The dynamic attack parameters are considered as $K^L_{4,1} = 6.8$, $K^L_{5,1} = 11.3$ and $K^L_{19,33} = 18$, for IEEE 6-,14-, 39- bus systems, respectively. The numerical results, which presents the $\eta_1$ and $\eta_2$ values, are shown in Table \ref{tbl:Accuracy_all_case}. The findings confirm the effectiveness of the SR method against other benchmarks in all cases. As shown in Table \ref{tbl:Accuracy_all_case}, the $\eta_1$ and $\eta_2$ values for SR method are in range $(0.028$ - $0.060)$ and $(0.011$ - $0.038)$, respectively, while larger error values are observed for other benchmark algorithms.}

\subsection{Multi-point Attacks}

{We also consider simulations with multi-point attacks (multiple victim buses). We only present results for the case of fast dynamics (due to the difficulties in training the other algorithms for slow-oscillatory dynamics, as mentioned in Section~5.2).} The attack parameters in this case are given by $K^L_{15,33} = 4, K^L_{19,33} = 14, K^L_{20,33} = 4.$ Once again, the parameters are chosen to represent a system that is stable without attacks, but unstable due to the LAAs. The accuracy results for the different attack parameters are plotted in Fig.~\ref{fig:Acc_MLAA} considering an observation time widow of $T = 16$ seconds. We observe that both SR and PINN algorithms produce reliable estimates within the considered observation time window. 

{As stated in section 3.3, the performance of the black-box machine learning algorithm such as SVM and NN relies on the range of the training data considered during the offline training phase. The simulation results also investigate this claim using two different case studies with different dynamic attack parameter ranges (i) $(1.2-23.4)$ and (ii) $(0.8-32.4)$. The outcome of the simulations is presented in Figs.~\ref{fig:Acc_MLAA_SVM_NN_30} and \ref{fig:Acc_MLAA_SVM_NN_80}. The results illustrate that by increasing the range of the dynamic attack parameters considered in the training data (Fig.~\ref{fig:Acc_MLAA_SVM_NN_80}),
the prediction accuracy decreases. This is expected since a lower range (of the attack parameters) in the training data set implies more points around the actual attack parameter used by the attacker, and hence, better prediction accuracy (conversely, lower accuracy for higher range). However, the operator has no way of having prior knowledge of the actual attack parameter that will be used by the attacker. Thus, they must consider a large range of the attack parameters in the training data set (otherwise they may risk misclassifying the attack completely).
Thus, black-box machine learning methods have a high dependency on the input data structure. However, in the proposed methods in this study, especially in SR, we do not have these of limitations, which confirms that SR can be a good solution for a wide range of system parameters.}

\begin{figure}[!t]
\centering
\includegraphics[trim=40 0 0 0,clip,width=0.55\textwidth]{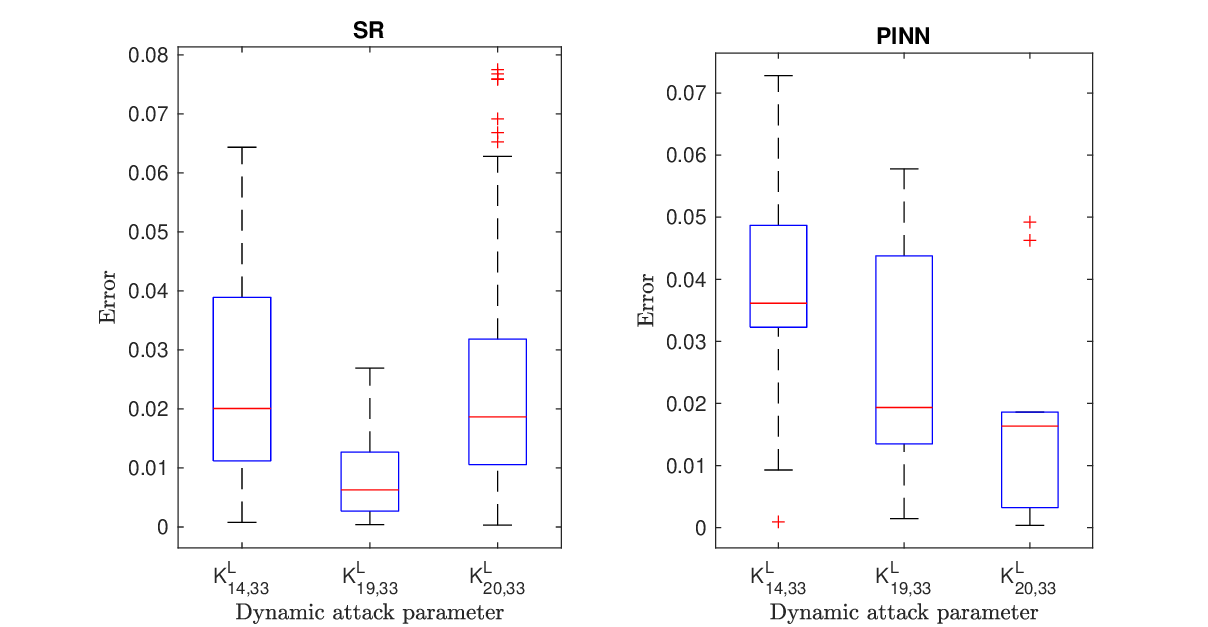}
\caption{Attack parameter estimation accuracy $\eta_1$ (boxplots) for multi-point attacks with T = 16 s. Attack parameters: $K^L_{14,33} = 4, K^L_{19,33} = 18, K^L_{20,33} = 4$.}
\label{fig:Acc_MLAA}
\end{figure}

\begin{figure}[!t]
\centering
\includegraphics[trim=40 0 0 0,clip, width=0.55\textwidth]{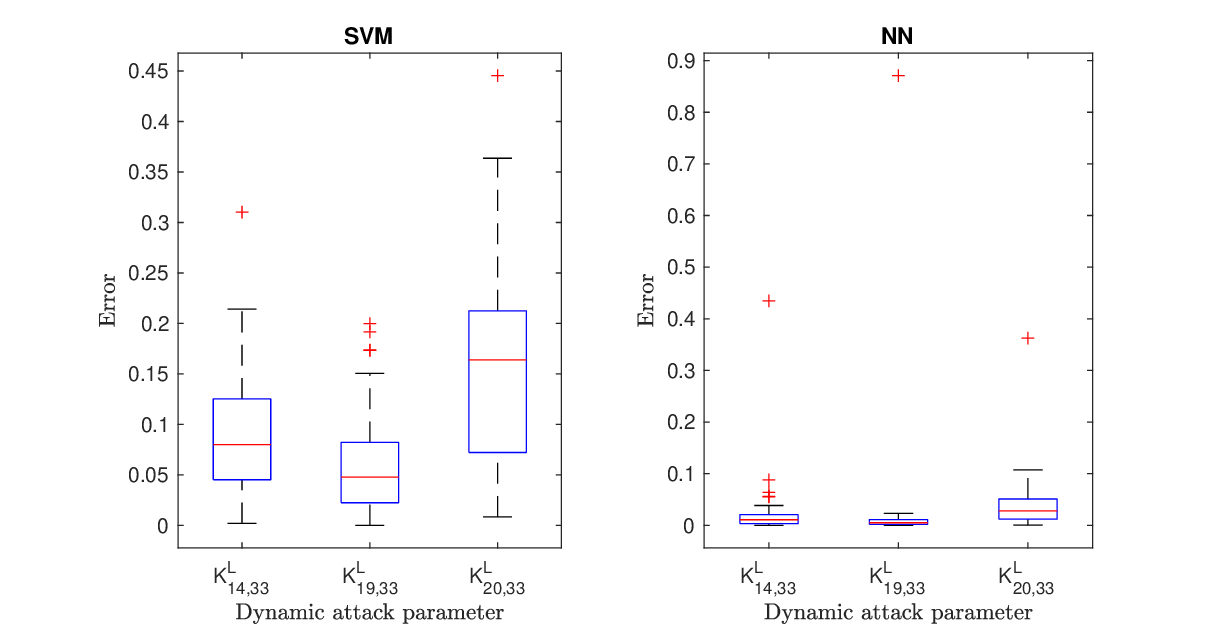}
\caption{{Attack parameter estimation error $\eta_1$ (boxplots) for multi-point attacks with T = 16 s for $K^L_{14,33} = 4 , K^L_{19,33} = 18, K^L_{20,33} = 4$. Attack parameters range considered in the offline-training procedure is $(1.2-23.4)$.}}
\label{fig:Acc_MLAA_SVM_NN_30}
\end{figure}

\begin{figure}[!t]
\centering
\includegraphics[trim=40 0 0 0,clip, width=0.55\textwidth]{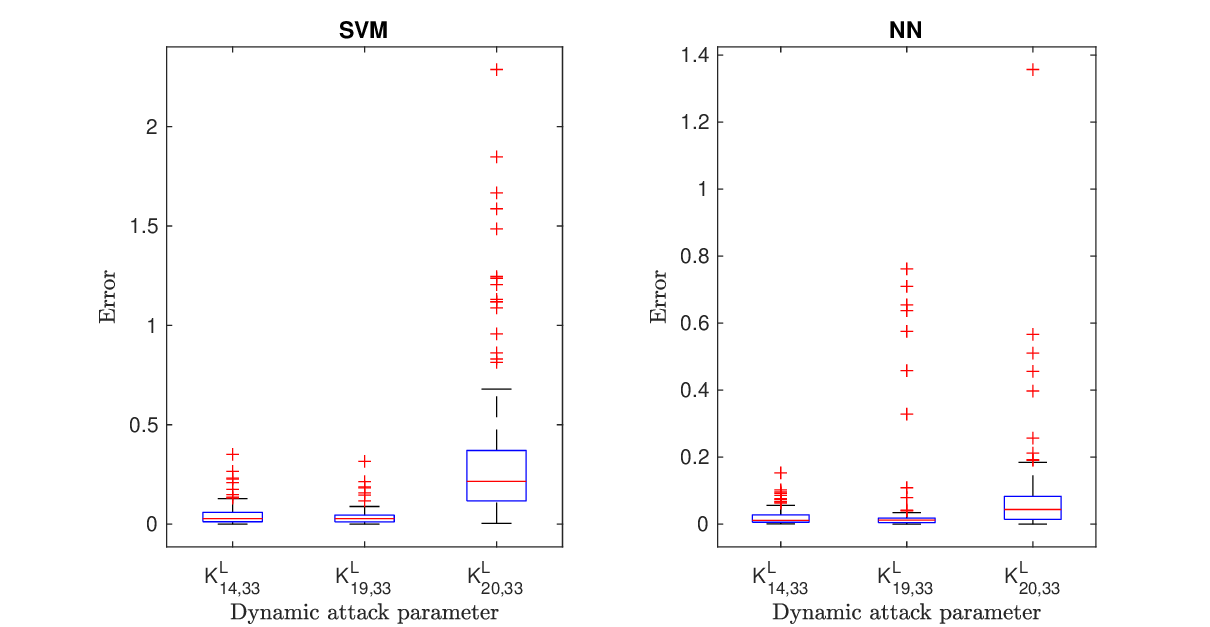}
\caption{{Attack parameter estimation error $\eta_1$ (boxplots) for multi-point attacks with T = 16 s for $K^L_{14,33} = 4 , K^L_{19,33} = 18, K^L_{20,33} = 4$. Attack parameters range considered in the offline-training procedure is $(0.8-32.4)$.}}
\label{fig:Acc_MLAA_SVM_NN_80}
\end{figure}

\subsection{Execution Times} 
{Finally, we also enlist the execution times of the SR, PINN, SVM, and NN algorithms.} The simulations are conducted on a Windows PC with Intel Xeon(R) CPU E5-2630 v3 @ 2.40GHz processor, RAM :24 GB GPU NVIDIA K80. For the PINN algorithm, we consider two training regimes. First, train from scratch, in which the NN weights are randomly initialized and trained with the observed data. In the second method, the NN weights are pre-trained with the phase angle/ frequency measurements with no attacks. Then during the online phase, the NN weights are initialized with the pre-trained weights and fine-tuned with the real-time measurements (with attack). The results are tabulated in Table~\ref{tbl:Exe_Time} averaged over $100$ runs. It can be noted that the SR algorithm takes significantly less time to execute compared to the PINN, algorithm. This is because the PINN involves training a NN over several epochs, which is computationally complex. The pre-training reduces the computational time of the PINN algorithm. On the other hand, SR follows the LASSO method which is computationally much simpler. {Thus, in addition to being broadly applicable over a wide range of system parameters, the SR method also has an additional advantage of quick execution in comparison with PINN, SVM, and NN algorithms, thus minimizing the system downtime.}

\begin{table}[!t]
\processtable{Execution times for the SR and PINN algorithms.\label{tbl:Exe_Time}}
 {\begin{tabular}{c  c  c  c  c  c } 
 \hline
 Method & Execution time (s)  \\ [0.5ex] 
 \hline\hline
SR & $4.84$   \\ \hline
PINN & $4071$  \\ \hline
{SVM} & {$27.13$}  \\ \hline
{NN} & {$69.15$}  \\
 \hline
\end{tabular}}{}
\end{table}

\section{Conclusions}
\label{sec:Conc}
{This study proposes data-driven algorithms to detect and identify IoT-enabled LAAs against power grids. To this end, a system identification problem is investigated, in which the attack parameters that best describe the observed dynamics are estimated from the measurements. Two physics-informed ML algorithms, namely, SR and PINN are proposed,  that estimate the attack parameters from the observed data. Their performance is tested extensively using different test systems, including IEEE 6-,14- and 39-bus systems. 
Other benchmark approaches, including SVM, NN and UKF, are also investigated to verify the effectiveness of the proposed methods. The numerical results confirm that the proposed data-driven algorithms outperform other benchmark techniques. SR method presents high precision in estimating dynamic attack parameters in single and multi-point attacks with about 3 \% error on average for fast and slow dynamics. However, the PINN algorithm does not perform well for systems with slow dynamics due to difficulties in training the neural network. The main advantage of SR and PINN compared to fully data-driven algorithms, such as SVM and NN, is their 
performance is not dependent on the offline training data structure. Future work includes (i) an extension to multi-area frequency control and optimal placement of PMUs to ensure reliable attack parameter estimation, (ii) machine learning algorithms to differentiate LAAs from generic faults in the system, and (iii) development of a co-simulation platform capable of emulating/simulating the cyber layer and simulating the physical layer (power grid).}

\section*{Appendix: Simulation Parameters}
System A1: Fast dynamics for IEEE 39-bus system
\begin{align*}
   & M_1 = 2.3186; 
   M_2 : M_8 = 2.6419;
    M_9 : M_{10}  = 2.4862. \\
    & K^P_1 - K^P_{10}= [1; 0.45; 0.45; 0.1; 0.5; 0.4; 0.3; 0.2; 0.4; 0.5]; \\
    & K^I_i = 0.6,  \forall i \in \mathcal{N}_G; D_i = 2,  \forall i \in \mathcal{N}_G; D_i = 0.01, \forall i \in \mathcal{N}_L; 
\end{align*}


System B1: Slow oscillatory dynamics for IEEE 39-bus system
\begin{align*}
    & M_1 = 2.3186; 
    M_2 : M_8 = 2.6419;
    M_9 : M_{10}  = 2.4862. \\
    & K^P_1 - K^P_{10}= [100; 45; 45; 10; 50; 40; 30; 20; 40; 50]; \\
    & K^I_i = 60,  \forall i \in \mathcal{N}_G; D_i = 2,  \forall i \in \mathcal{N}_G;  D_i = 0.01, \forall i \in \mathcal{N}_L; 
\end{align*}

{
System A2: Fast dynamics for IEEE 14-bus system
\begin{align*}
    & M_1 - M_5 = [0.125; 0.034; 0.016; 0.010; 0.015]; \\
    & D_1 - D_5 = [0.125; 0.068; 0.032; 0.068; 0.072]; \\
    & K^P_1 - K^P_{5}= [0.02; 0.09; 0.03; 0.03; 0.08]; \\
    & K^I_1 - K^I_5  =[0.35; 0.40; 0.35; 0.35; 0.40];\\
    &D_i = 0.01, \forall i \in \mathcal{N}_L; 
\end{align*}}

{
System B2: Slow oscillatory dynamics for IEEE 14-bus system
\begin{align*}
    & M_1 - M_5 = [0.125; 0.034; 0.016; 0.010; 0.015]; \\
    & D_1 - D_5 = [0.125; 0.068; 0.032; 0.068; 0.072]; \\
    & K^P_1 - K^P_{5}= [2; 9; 3; 3; 8]; \\
    & K^I_1 - K^I_5  =[35; 40; 35; 35; 40];\\
    &D_i = 0.01, \forall i \in \mathcal{N}_L; 
\end{align*}}

{
System A3: Fast dynamics for IEEE 6-bus system
\begin{align*}
    & M_1 - M_3 = [1.25; 1.25; 1.25;]; \\
    & D_1 - D_3 = [0.125; 0.125; 0.125;]; \\
    & K^P_1 - K^P_{5}= [0.02; 0.09; 0.03;]; \\
    & K^I_1 - K^I_3  =[0.35; 0.40; 0.35;];\\
    &D_i = 0.01, \forall i \in \mathcal{N}_L; 
\end{align*}}

{
System B3: Slow oscillatory dynamics for IEEE 6-bus system
\begin{align*}
    & M_1 - M_3 = [1.25; 1.25; 1.25;]; \\
    & D_1 - D_3 = [0.125; 0.125; 0.125;]; \\
    & K^P_1 - K^P_{5}= [2; 9; 3;]; \\
    & K^I_1 - K^I_5  =[35; 40; 35;];\\
    &D_i = 0.01, \forall i \in \mathcal{N}_L; 
\end{align*}}

\bibliographystyle{IEEEtran}
\bibliography{bibliography}

\end{document}